\documentclass[draft]{llncs}

\usepackage{graphicx}
\usepackage[T1]{fontenc}

\begin{document}
\pagestyle{empty}
\title{Size of the stable population in the Penna bit-string model of biological aging}
\author{K.Malarz \and M.Sitarz \and P.Gronek \and A.Dydejczyk}
\institute{Department of Applied Computer Science,
Faculty of Physics and Nuclear Techniques,
AGH University of Science and Technology\\
al. Mickiewicza 30, PL-30059 Krak\'ow, Poland\\
\email{malarz@agh.edu.pl}
}
\maketitle
\begin{abstract}
In this paper the Penna model is reconsidered. 
With computer simulations we check how the control parameters of the model influence the size of the stable population.
\end{abstract}

\section{Introduction}
The Penna model \cite{pennamodel} is popular and efficient way to simulate biological aging in genetically heterogeneous population (see \cite{stauffer00,mossdeoliviera99,mossdeoliveira98,bernardes96} for review).
It is based on {\em the mutation accumulation} theory which claims that random hereditary deleterious mutations accumulate over several generations in the species genome \cite{partidge93}.

The model is governed by a set of parameters such as environmental capacity $N_{max}$ so that the actual population $N\le N_{max}$.
The birth rate $B$ is the number of offsprings from a single parent per evolution time step, providing the parent reached age $R$ and is still below maximum reproduction age $E$.
Other parameters which control the growth of the population are: number of mutations $M$ injected into the baby's genome, apart from the already inherited, and threshold value $T$ of {\em active} bad mutations above which items are eliminated for this only reason.

	In this paper we would like to check how model control parameters (described in the next section) influence the size $N(t)$ of the stable population for long times $t\to\infty$ when {\em mutational meltdown} \cite{lynch90,bernardes95,penna95,pal96,feshenfeld03,puhl95} is avoided.

\section{Model}
In the Penna model \cite{pennamodel} each individual in the population is characterised by its genome --- the $N_{bit}$-long binary string.
Time is measured by a discrete variable $t$.
In each time step ($t\to t+1$), for each individual of age $a$, the number of bits set to one in the first $a$ positions $i$ in genome is calculated.
Bit `1' represents a bad mutation activated when its position $1\le i\le a$, however mutations above $a<i\le N_{bit}$ are not harmful.
If the number of {\em bad} and {\em active} mutations is greater or equal to the threshold value $T$ an individual dies due to too many deleterious mutations and so called {\em genetic death} takes place.

Individuals competite among themselves for food and territory: each of them may be removed from the population with probability $N(t)/N_{max}$, where $N_{max}$ represents the maximal environmental capacity.
In other words, we introduce the Verhulst's factor in order to avoid exponential growth of the population size to infinity.
The number $N_{bit}$ of bits in genome restricts also maximal age of individuals.

We start the simulation with $N_{ini}$ individuals which genomes contain only `0' (no bad mutations).
The population reproduces asexually.
If the individual is older than minimum reproduction age $R$ it is able with the probability $b$ to give $B$ offsprings.
The offspring's genome is a copy of parent's one, exposed during replication to $M$ harmful mutations.
Each mutation occurs with probability $m$ at randomly chosen position in the genome.
We also introduce a maximum reproduction age $E$, so the individual older than $E$ do not clone itself any more.

\section{Results}
\begin{figure}
\begin{center}
(a) \includegraphics[bb=0 0 595 842,angle=-90,width=.45\textwidth]{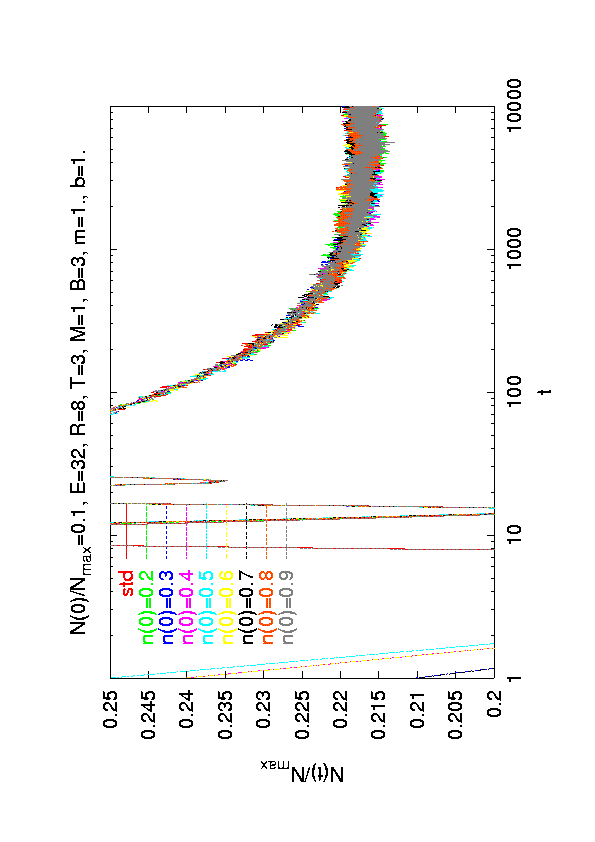}
(b) \includegraphics[bb=0 0 595 842,angle=-90,width=.45\textwidth]{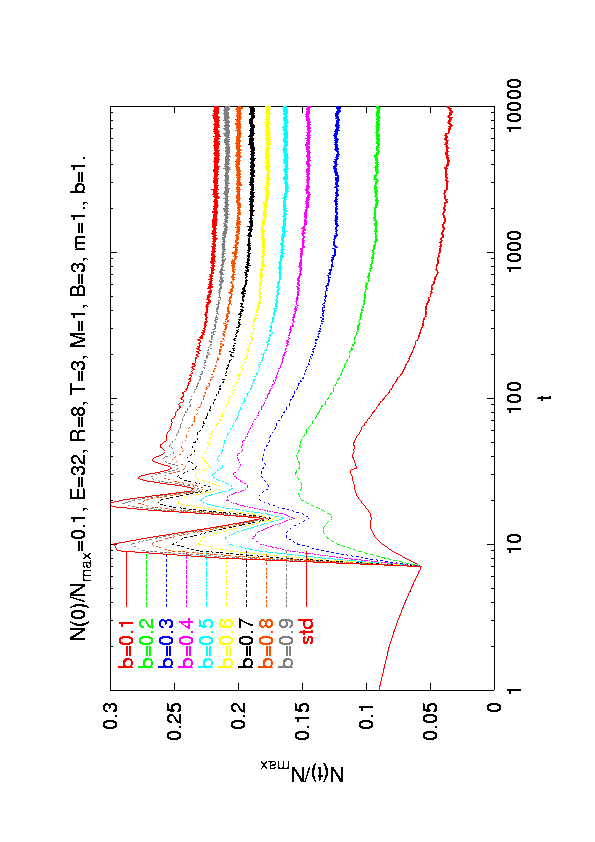} \\
(c) \includegraphics[bb=0 0 595 842,angle=-90,width=.45\textwidth]{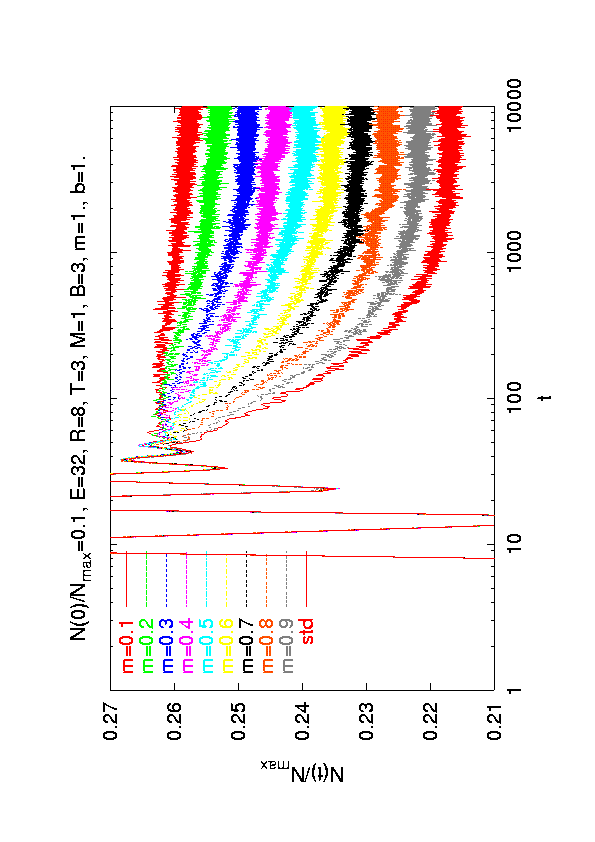}
(d) \includegraphics[bb=0 0 595 842,angle=-90,width=.45\textwidth]{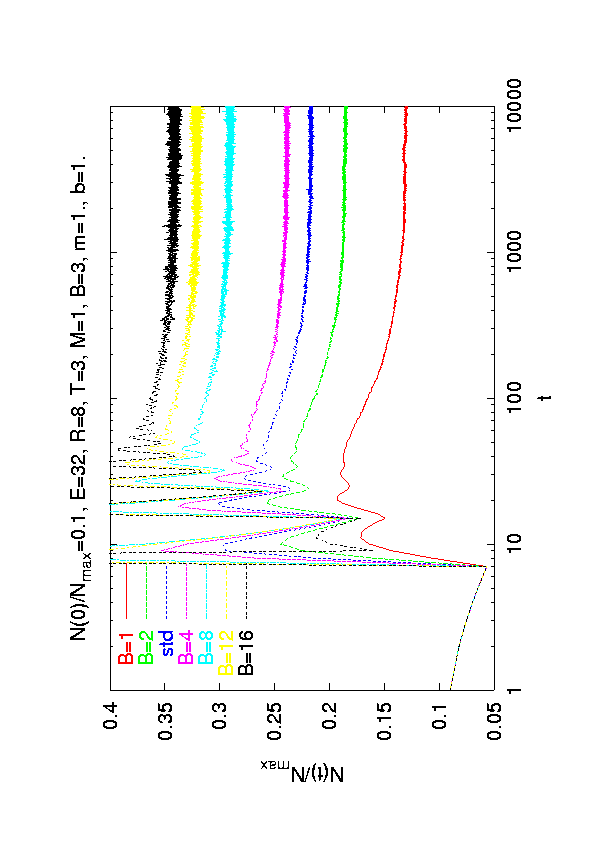} \\
(e) \includegraphics[bb=0 0 595 842,angle=-90,width=.45\textwidth]{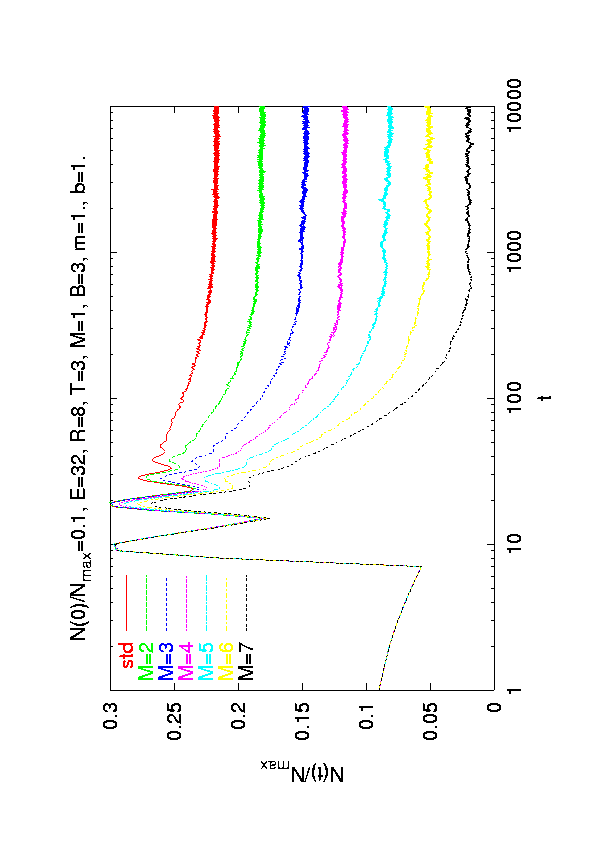}
(f) \includegraphics[bb=0 0 595 842,angle=-90,width=.45\textwidth]{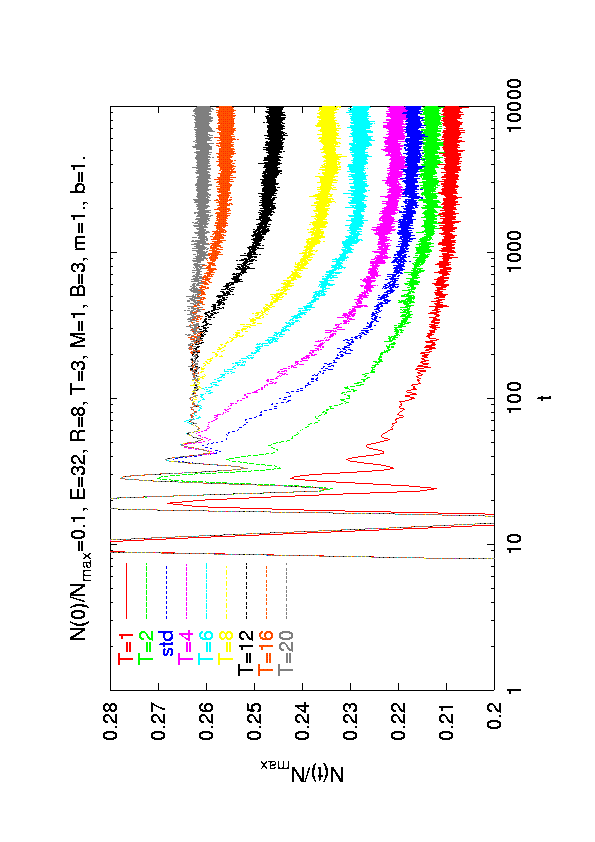} \\
(g) \includegraphics[bb=0 0 595 842,angle=-90,width=.45\textwidth]{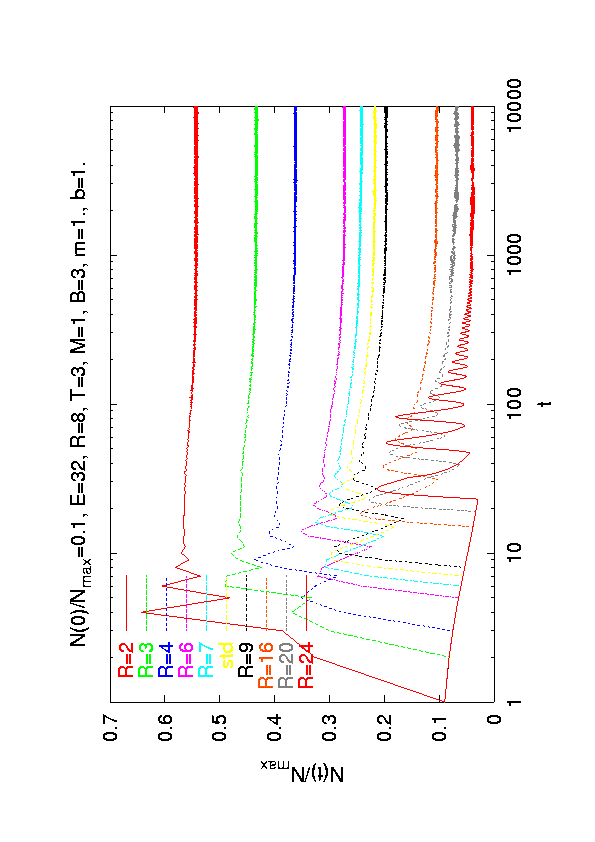}
(h) \includegraphics[bb=0 0 595 842,angle=-90,width=.45\textwidth]{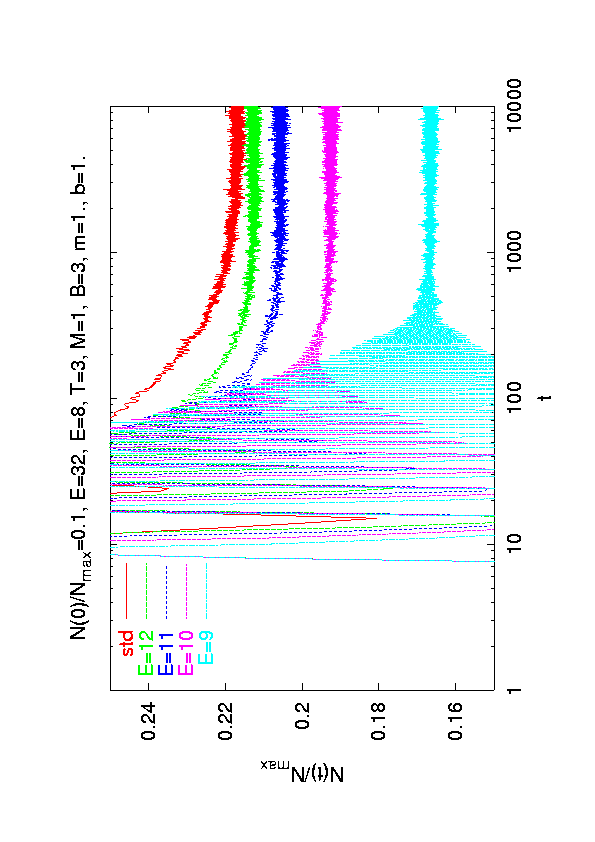}
\caption{The influence of the model control parameter (a) $n(0)$, (b) $b$, (c) $m$, (d) $B$, (e) $M$, (f) $T$, (g) $R$, and (h) $E$ on the population size.}
\label{fig}
\end{center}
\end{figure}

We would like to refer our results to the standard one (labelled as `std' in all figures) which we assume to be for $N_{bit}=32$, $N_{max}=10^6$, $n(0)=N(0)/N_{max}=0.1$, $R=8$, $E=32$, $T=3$, $M=1$, $B=3$, $m=1.0$ and $b=1.0$.
The influence of the $N_{bit}$ on population characteristics was discussed earlier in \cite{penna96,malarz00}.
The maximal environment capacity $N_{max}$ is fixed during simulation.
The initial concentration of individuals in respect to maximal environmental capacity do not influence the results (Fig.~\ref{fig}(a)).

As one may expect the larger probability of reproduction $0\le b\le 1$ gives the larger population size (Fig.~\ref{fig}(b)).
The population size decreases when the probability of mutations $0\le m\le 1$ increases (Fig.~\ref{fig}(c)).

The same situation is for integer variables $B$ and $M$ (see Figs. \ref{fig}(d) and \ref{fig}(e), respectively).
If the number of mutations is too large (i.e. $M>7$) the mutational meltdown occurs.
Also too large value of the number of offsprings (e.g. $B\ge 20$) may lead to catastrophe: too rapid growth of the population may cause that in one time step the environmental maximal capacity $N_{max}$ will be exceeded.
As a consequence of that the population will vanish {\em only} due to Verhulst's factor and not the genetic one.
In our implementation products $Mm$ and $Bb$ may be considered as the average number of mutations per individual's genome and as the average number of offsprings given every year per reproductive individual (i.e. in age $R\le a\le E$), respectively.

In Fig.~\ref{fig}(f) the influence of the threshold $T$ on the population size is presented.
Decreasing the threshold $T$ results in increasing of the stable population size.
Setting $T=N_{bit}$ eliminates genetic death and again only Verhulst's factor determinates $N(t\to\infty)$.

Delay in starting the reproduction decreases the population size as presented in Fig.~\ref{fig}(g).
Obviously, setting $R>N_{bit}$ destroys population because individuals never replicate themselves.
For $R=1$ --- when all newly born babies become adults as quickly as possible --- a very strong oscillations appear (not shown).

Fig.~\ref{fig}(h) shows the influence of the maximal reproduction age $E$ on the stable population size.
For $E=8(=R)$, when individual reproduce {\em only once} (semelparous organisms \cite{penna95-2}) oscillations in the population size appears (not shown).

\section{Conclusions}
The population size of the stable population in the Penna model was already investigated for a certain purpose, e.g.: 
showing advantage of sexual reproduction on asexual one \cite{bernardes97,martins01,stauffer99,stauffer01},
studying sexual fidelity of males vs higher reproduction rate \cite{sousa99},
modelling fishing and/or hunting \cite{mossdeoliveira95-2,cebrat97,feingold96},
explaining mystery of Pacific salmon senescence \cite{penna95-2},
analysing pray-predator systems \cite{puhl95},
recovering demographic data \cite{penna96,stauffer99,sitarz03},
modelling the oldest old effect \cite{mossdeoliveira95},
modelling consequence of parental care \cite{mossdeoliveira99,sousa99-2},
studying the individuals fertility \cite{argollodemenezes96},
studying the migration effect \cite{magdon99},
and others \cite{maksymowicz99,maksymowicz99-2,makowiec97,laszkiewicz02,vandewalle99,penna95-3,ito96,dealmeida98,dealmeida98-2,sousa00,magdon02}.

In this paper we confine our attention to the basic features of the model.
Main results were collected in Figs. \ref{fig}.
We may conclude the following:
\begin{enumerate}
\item Different time scales must be accounted for, if the `stable' refers to different properties observed in the computer experiment;
  \begin{itemize}
  \item total population $N(t)$ shows values that do not change significantly after about one hundred iteration steps,
  \item  however, we need at least a couple of thousand steps for age distribution \cite{malarz00,mossdeoliveira95} and, especially, bad mutations distributions in a genome position \cite{puhl95,bernardes97}, to become stationary.
  \end{itemize}
\item Obviously for smaller birth rates $Bb$ the population is smaller.
However, unlike the logistic model for which $N(t\to\infty)=BbN_{max}/(1+Bb)$ \cite{maksymowicz99,magdon02}, the Penna model shows a critical $B_C$ below which the population is extinct.
This is due to the genetic death as result of pumping in more  and more mutations $M$ into each new generation, and so $B_C(M)$ with the limit $B_C=0$ for $M=0$, the logistic case.
\item When the number of bad mutations $M$ injected into baby's genome is scaned from $M=0$ (the logistic model), the upper limit $M_C$ for the extinction strongly depends on fertility:
  \begin{itemize}
  \item if $B$ is smaller, $M_C$ is also smaller, and
  \item for sufficiently large $B$, $M_C\to\infty$ as the high birth rate gives a chance to new population (however rich in bad mutations) to reproduce before it dies out soon on reaching the threshold $T$.
	This is so if the minimum reproduction age $R$ is sufficiently small to allow for the reproduction.
  \end{itemize}
\item The upper age limit $E$ for reproduction influence the stable population $N$ and larger $E$ makes bigger $N$, as it may be expected.
This time, however, we do not observe any critical behaviour when we scan $E$.
\end{enumerate}

\section*{Acknowledgements}
The authors are grateful to Prof. A.~Z.~Maksymowicz for fruitful discusion.
Simulations were carried out in ACK-CYFRONET-AGH.
The machine time on SGI~2800 is financed by the Ministry of Scientific Research and Information Technology in Poland under grant No. KBN/\-SGI-\-ORIGIN-\-2000/\-AGH/\-094/\-1999.



\end{document}